\documentclass[prl,twocolumn,amsmath,amssymb,showpacs,superscriptaddress,floatfix]{revtex4}

\usepackage{graphicx}
\usepackage{float}
\usepackage{url}
\usepackage{color}
\newcommand{\celc}[0]{^\circ\mathrm{C}}

\usepackage{graphicx}
\begin{document}
\title{Measurement of correlations between low-frequency vibrational modes and particle rearrangements in quasi-two-dimensional  colloidal glasses}

\author{Ke Chen}
\affiliation{Department of Physics and Astronomy, University of Pennsylvania, Philadelphia, Pennsylvania 19104, USA}

\author{M. L. Manning}
\affiliation{Princeton Center for Theoretical Science, Princeton University, New Jersey, 08544, USA}

\author{Peter J. Yunker}
\affiliation{Department of Physics and Astronomy, University of Pennsylvania, Philadelphia, Pennsylvania 19104, USA}

\author{Wouter G. Ellenbroek}
\affiliation{Department of Physics and Astronomy, University of Pennsylvania, Philadelphia, Pennsylvania 19104, USA}

\author{Zexin Zhang}
\affiliation{Department of Physics and Astronomy, University of Pennsylvania, Philadelphia, Pennsylvania 19104, USA}
\affiliation{Complex Assemblies of Soft Matter, CNRS-Rhodia-UPenn UMI 3254, Bristol, Pennsylvania 19007, USA}
\affiliation{Center for Soft Condensed Matter Physics and Interdisciplinary Research, Soochow University, Suzhou 215006, China}

\author{Andrea J. Liu}
\affiliation{Department of Physics and Astronomy, University of Pennsylvania, Philadelphia, Pennsylvania 19104, USA}

\author{A. G. Yodh}
\affiliation{Department of Physics and Astronomy, University of Pennsylvania, Philadelphia, Pennsylvania 19104, USA}

\date{\today}
\begin{abstract}
We investigate correlations between low-frequency vibrational modes and rearrangements in two-dimensional colloidal glasses composed of thermosensitive microgel particles which readily permit variation of sample packing fraction. At each packing fraction, the particle displacement covariance matrix is measured and used to extract the vibrational spectrum of the ``shadow'' colloidal glass (i.e., the particle network with the same geometry and interactions as the sample colloid but absent damping). Rearrangements are induced by successive, small reductions in packing fraction. The experimental results suggest that low-frequency quasi-localized phonon modes in colloidal glasses, i.e., modes that present low energy barriers for system rearrangements, are spatially correlated with rearrangements in this thermal system.
\end{abstract}
\pacs{82.70.Dd, 63.50.Lm}
\maketitle

In crystalline solids, plastic deformation is well understood to be associated with structural defects, particularly dislocations \cite{merminbook}. In amorphous solids, however, geometric structural parameters are not obviously connected to regions that later rearrange \cite{harrowellJPC}, hampering \emph{a priori} identification of fragile regions.  Recent numerical simulations suggest that the spatial distribution of low frequency phonon modes may be correlated with irreversible rearrangements in glasses \cite{harrowellNP, brito2010}, and that the quasi-localized low frequency vibrational modes often observed in glasses \cite{andersonbook}  play a role in glass mechanical response \cite{tsamados2009,tanguy2010,xu2010}.  In a related vein, experiments in vibrated granular packings find that when cracks begin to appear, particles are likely to move in the direction of the polarization vectors of the lowest frequency modes \cite{brito2010}. Taken together, these results suggest connections between rearrangement events and low-frequency vibrational modes in some glassy systems.  They raise the question of whether such connections also exist in atomic and molecular glasses..

In this paper, we experimentally investigate correlations between vibrational modes and particle rearrangements in quasi-two-dimensional (quasi-2D) colloidal glasses. The work is made possible by recently developed covariance matrix techniques that have led to the measurement of vibrational modes in thermal colloids from microscopy experiments \cite{ghosh2010, kaya2010, chen2010}. The colloidal glasses are composed of thermosensitive microgel particles which readily permit \emph {in situ} variation of sample packing fraction.  At each packing fraction, the particle displacement covariance matrix of the sample is measured and used to extract the vibration spectrum of the ``shadow" colloidal glass, i.e., the particle network with the same geometry and interactions as the sample colloid but absent damping.  Particle rearrangements are observed to occur when the packing fraction, or compressive stress, is decreased slightly.  The experiments thus permit comparison of packing-fraction-induced rearranging particle clusters with the spatial distribution of glassy phonon modes. Note that if our highly damped system were athermal, then rearrangements would only occur when an energy barrier--and, correspondingly, the frequency of a related vibrational mode--vanishes.  In that case, the rearrangements would be deterministic. Our thermal system, however, can surmount low energy barriers associated with modes at nonzero frequency, leading to a stochastic response.  Nonetheless, we find spatial correlations between regions of high displacement in low frequency quasi-localized phonon modes (``soft spots"~\cite{manning}) and rearranging clusters of particles in glasses.  Thus, our experiments illustrate that soft spots are correlated with rearrangements even when thermal fluctuations are present.  Such resilience is critical if soft spots are to be viewed as the ``structural defects" in disordered systems, analogous to dislocations in crystals, that control mechanical response to applied load. 
  
The samples are disordered quasi-2D glasses composed of poly(N-isopropyl acrylamide) microgel colloidal spheres (i.e. NIPA particles) suspended in water. The NIPA particle diameter increases when temperature is reduced \cite{vincent1999, pelton}.  Therefore, the sample packing fraction can be tuned over a relatively wide range, with suspension evolving from colloidal fluid to jammed solid, via temperature changes of just a few degrees. The inter-particle potentials are measured to be short-range repulsive with a soft tail \cite{zhang2009}, enabling particles to be packed at high concentration.  A binary mixture of microgel particles, with diameters of 1 $\mu$m and 1.4 $\mu$m and with a large/small particle number ratio of 0.7, was loaded between two glass cover slips to create a colloidal monolayer. The sample was hermetically sealed using optical glue (Norland 63) and annealed for 2 hours at 28 $\celc$.   Images of the samples were acquired at 30 frames/s using bright field video microscopy at temperatures ranging from 24.9 to 26.3 $\celc$. The sample temperature controls the proximity of the sample packing fraction to the jamming point, which is determined by the maximum of the height of the first peak in the pair correlation function to be $\phi_{J}$ = 0.84 \cite{zhang2009}. 
 When the temperature is increased slightly ($\sim$0.2 $\celc$), the particles shrink in size in less than 0.1 s\cite{yunker2009}, and they rearrange to find new equilibrium positions \cite{supp} as a result of unbalanced stresses in the system. A small decrease in particle size thus induces a uniform and unbiased perturbation of the system, similar to thermal annealing in bulk atomic and molecular materials. Sample temperature was controlled by thermal coupling to a resistively heated microscope objective (BiOptechs) and the sample was permitted to equilibrate for 15 min at each temperature before data acquisition. The observation field-of-view was far from any boundaries or non-uniform regions within the sample, and trajectories of particles in the field-of-view were extracted using standard particle tracking techniques \cite{crocker1996}. Cage rearrangements were not observed during the 1000 second runs at constant packing fraction. Thus, the system remained in the same basin of the energy landscape. Brief rearrangement events were observed as packing fraction was reduced, until the system settled down into another well-defined average configuration. 

\begin{figure}[!t]
\includegraphics[width=8cm]{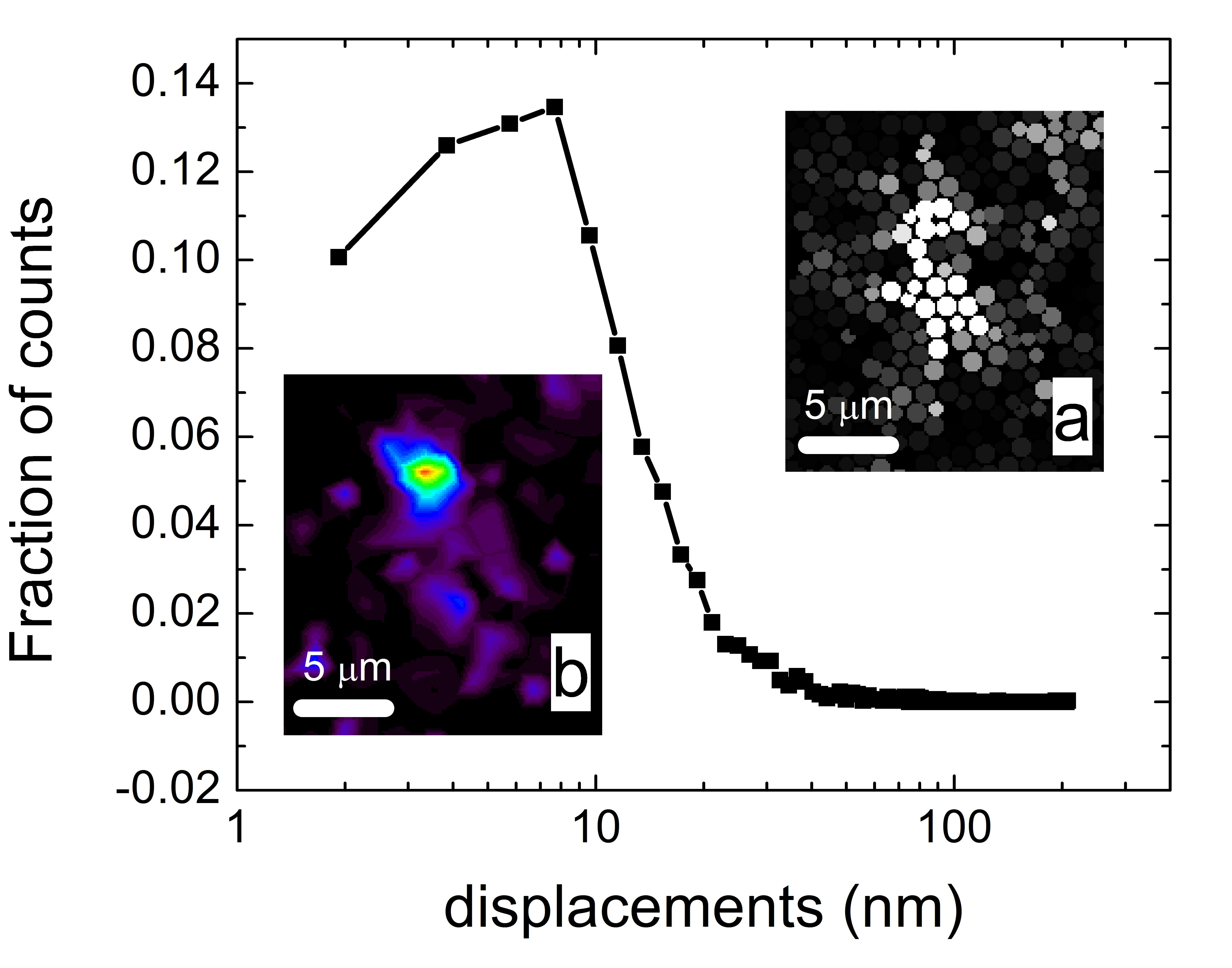}

\caption{(color online) Particle displacement distribution when packing fraction changes from 0.885 to 0.879. Inset a: real space distribution of particle displacements; brightness of a particles is proportional to the displacement magnitude. Inset b: real space distribution of change of $\Psi_{6}$  of the same region as Inset a, color change from black to red represents a change of $\Psi_{6}$ from 0 to 0.89, respectively. }

\label{fig:displ}
\end{figure}  

The particle rearrangements by compression are spatially heterogeneous: some particles move, often in clusters, but most particles do not move measurable distances.  We resolve particle positions with $7$ nm spatial resolution.  The distribution of particle displacements when the packing fraction $\phi$ was changed from $\phi$ = 0.885 to $\phi$ = 0.879 is plotted in Fig.~1. Notice that most particles experience a displacement of less than 40 nm, i.e., less than~1/30 of a particle diameter. In our analysis, these particles are considered to have remained in their equilibrium position during the temperature change; this non-zero displacement is due to the drifting of equilibrium positions as a result of elastic relaxation, and the diffusion of particles within cages over long periods of time ($\sim$2000 seconds). Such drifting is typically less than 40 nm, and it does not amount to a structural rearrangement.  These drifts, therefore, do not influence the structural or dynamical relation between particles, since the particles remain in the same cages of their neighbors. The distribution has a long tail due to particles which have experienced significant displacements. The latter particles tend to be spatially clustered as shown in Fig.~1 (Inset a). We also measure the change in the local bond orientational order parameter, i.e., for the \emph{j}th particle, $\Psi_{6}$ =$\frac{1}{N_{nn}}$$\sum_{k}^{N_{nn}}{e^{6i\theta_{jk}}}$, summed over all nearest neighbor bonds; $\theta_{jk}$ is the angle between the bond joining particles \emph{j}, \emph{k} and the x-axis, and $N_{nn}$ is the number of nearest neighbors. In Inset b of Fig.1, the distribution of the absolute value of each particle's change in bond orientational order, is plotted. This peak in the change of bond orientational order is spatially correlated with clusters of particles with large displacement.  Thus, both particle displacements and changes in bond orientational order demonstrate the heterogeneous nature of rearrangements in the packing.

\begin{figure*}[!t]
\includegraphics[width=18 cm]{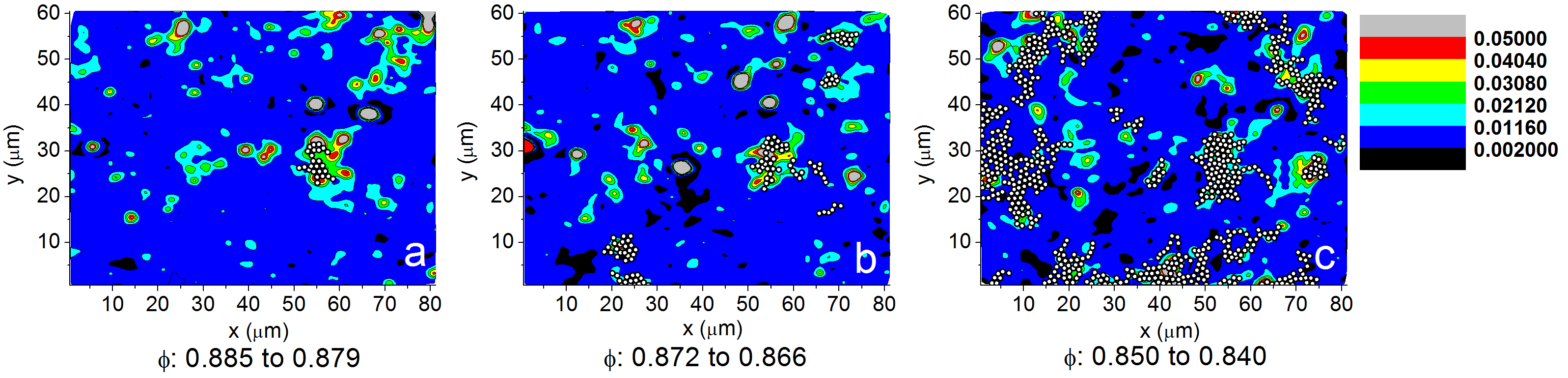}
\caption{(color online) Rearranging clusters and low-frequency modes at three different packing fractions studied, ranging from $\phi=0.885$ to $\phi=0.850$ from a to c.  Color contour plots indicate polarization magnitudes for each particle, summed over the lowest 30 modes. Circles indicate particles that moved more than 40 nm at each decompression step.  
}
\label{fig:cont}
\end{figure*}  

The measured displacement covariance matrix \cite{ghosh2010, kaya2010, chen2010, yunker2011} was employed to extract vibrational modes of the ``shadow'' colloidal system, which shares the same geometric configuration and interactions of the experimental colloidal system, but is undamped. A displacement covariance matrix is constructed by calculating the equal-time covariance of displacements between each particle pair for each direction. The relative amplitude and direction of motion due to a mode $\omega$ is thus described in terms of the polarization vector $\vec e_{i}^{\omega}$ for each particle $i$.

Fig.~2 exhibits colored contour plots of polarization magnitudes for each particle summed over the 30 lowest frequency modes, i.e., $\frac{1}{30}$$\sum_{\omega_{1}}^{\omega_{30}}({e_{i\gamma}^{\omega})^2}$, where $e_{i\gamma}^{\omega}$  is the $\gamma$ coordinate of the polarization vector $\vec e_i$ of particle \emph{i} for a normalized eigenmode with frequency $\omega$, as in Ref.~3. These vibrational properties correspond to the packing configuration before the packing fraction change. In addition, the circles plotted in the figure correspond to particles that experienced significant displacements (i.e., displacements larger than 40 nm) following the packing fraction change. Use of the change in $\Psi_{6}$ gives a similar selection of clusters. Since we are most interested in large rearrangement events, we group these displaced particles into clusters based on nearest neighbor pairings, with clusters containing fewer than five particles ignored. A good correlation between the spatial distribution of the low frequency modes and the rearranging clusters is apparent over several steps of change in packing fraction \cite{supp}. (Note, these results are not sensitive to the number of modes chosen; similar results are obtained using the lowest 10 or 50 modes.)  We see that the number of rearranging clusters, as well as the total number of rearranging particles, increases as $\phi$ is decreased towards the jamming transition as the system gradually loses its mechanical stability. Note that participation in low frequency modes correlates with regions of substantial particle rearrangement, as found in Ref.~3.  However, rearrangements studied in Ref.~3 only involved particles that irreversibly changed four nearest neighbors (or more), and involved samples below the jamming point. In our experiments, no particle experienced more than four nearest neighbor changes. 

The correlation between the vibrational modes and rearrangements can be quantified by the normalized inner product between the eigenvectors, which forms a complete basis for 2N-component vectors \cite{brito2010}, and the displacement vector $\vec{R}_{d}$ that contains the displacement components of all particles in the field of view. Here N is the number of particles in the field of view. A projection coefficient is defined as $\alpha^2(\omega)$ = $\frac{|\vec{e}^\omega\cdot \vec{R}_{d}|^2}{|\vec{R}^2_{d}|}$, where $\vec{e}^\omega$ is the eigenvector of mode $\omega$.  $\alpha^2(\omega)$ for all packing fraction steps is plotted in Fig.~3. Clearly, the contributions of the lowest frequency modes (particularly the lowest $N_m \approx 10$ modes) to the rearrangement vectors are much larger than those from higher frequency modes, and they are more than two orders of magnitude larger than the correlation noise of 1/2N.  Furthermore, the accumulated projection coefficients of the lowest frequency modes are significant  \cite{supp}. Thus, rearrangements are not only likely to occur in peak regions of low frequency modes, but the displacements of individual particles also tend to follow the directions of the polarization vectors of the these modes, consistent with numerical findings for sheared systems \cite{tanguy2010}. This experimental finding strongly supports the notion that the quasi-localized low frequency modes present the lowest energy barriers to collective motion \cite{xu2010}. 

\begin{figure}[!t]
\includegraphics[width=8cm]{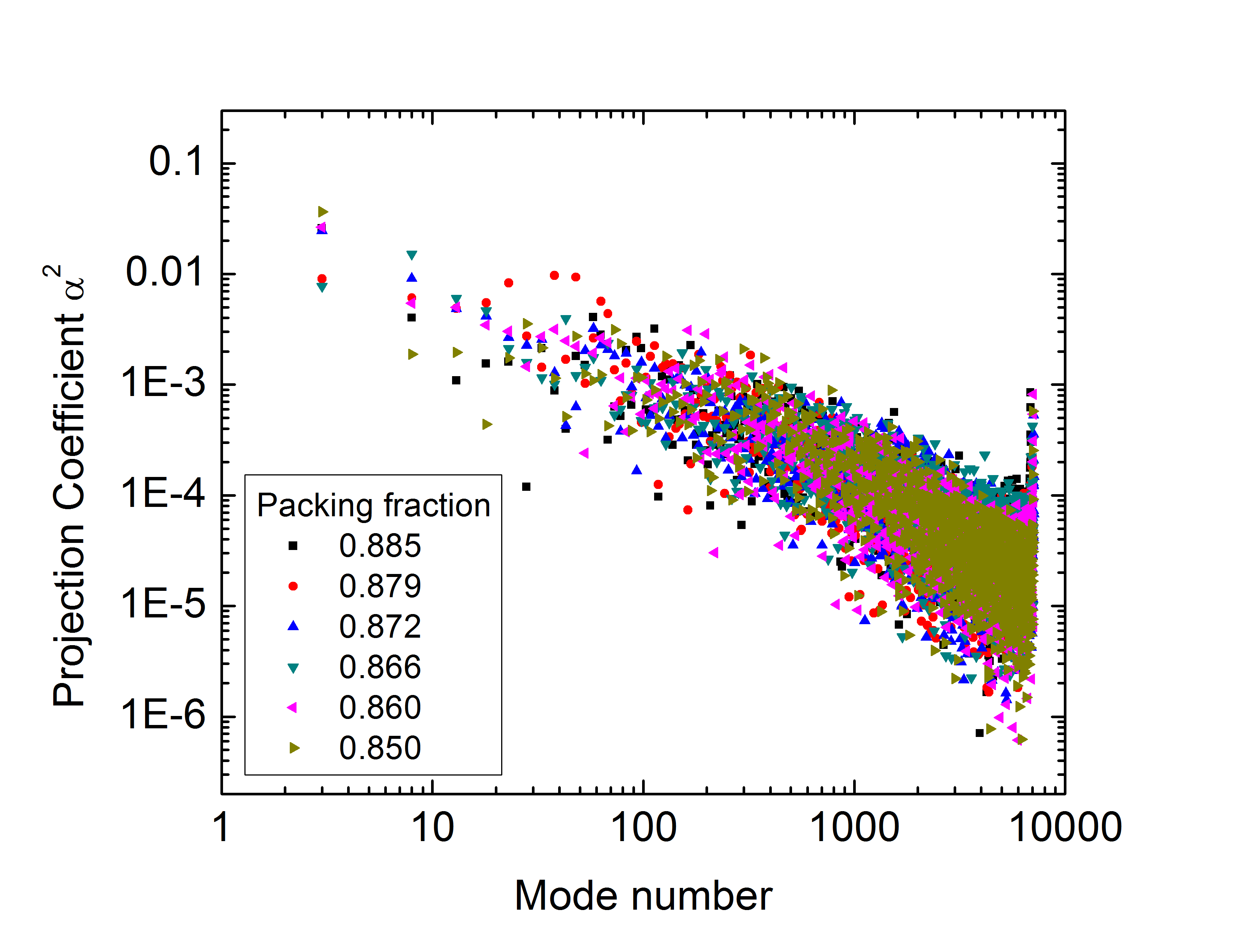}

\caption{(color online) Projection coefficient $\alpha^2$ versus mode number. Different symbols represent $\alpha^2$ from different packing fraction steps. The eigenvectors are derived at packing fractions as indicated by the symbols, and the rearrangement vector is determined from the displacements of particles when the system is decompressed from the packing fraction indicated to the next highest value. The data are binned with a window of 5.}

\label{fig:proj}
\end{figure}  

\begin{figure*}[!t]
\includegraphics[width=15 cm]{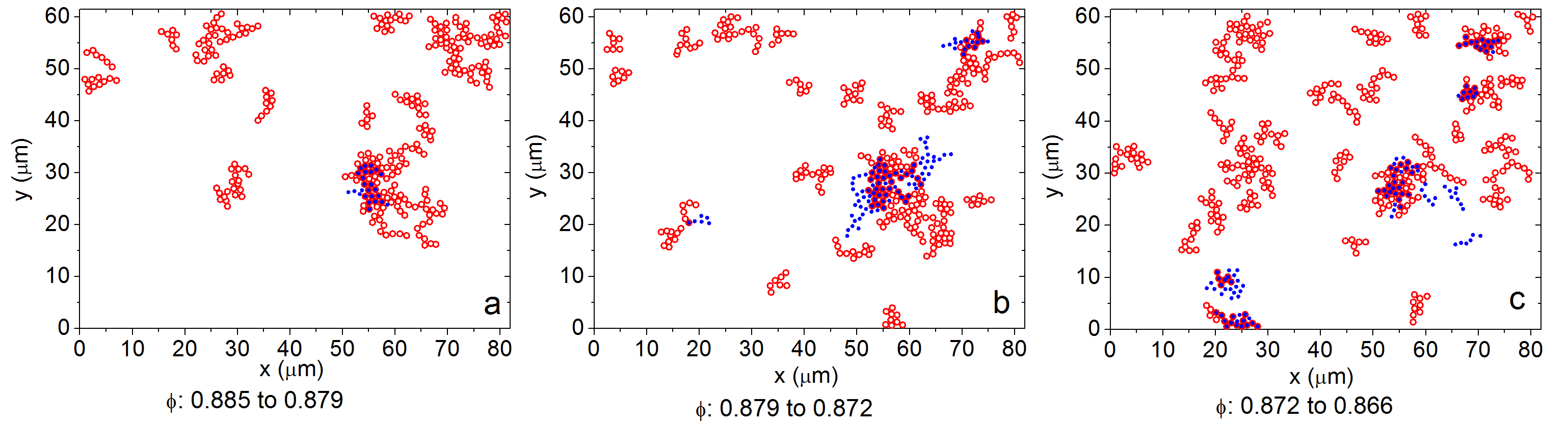}

\caption{(color online) Comparison of soft spots and rearranging clusters at high packing fractions.  Blue solid circles indicate particles belonging to rearranging clusters, defined as particles with displacements exceeding $40$nm; red empty circles are particles belonging to soft spots for $\emph{N}_{m}$  = 10, $\emph{N}_{p}$ = 130. }
\label{fig:soft}
\end{figure*} 

Thus far, our experiments suggest that rearrangements in disordered colloidal packings tend to occur in regions where low frequency phonon modes concentrate. However, not all such regions experience rearrangements, as shown in the high packing fraction results in Fig.~2a and b. Recent numerical work on athermal packings suggests that the mechanical response may be controlled by a population of soft spots, consistent with phenomenological theories of shear transformation zones \cite{langerstz}, where at least one soft spot is correlated to a given rearrangement \cite{manning}.  We now show that soft spots also exist in our colloidal system at finite temperatures. 


Simulations in Ref [12] show that regions prone to rearrangement or ``soft spots" are characterized by an important energy scale and length scale. The energy scale is characterized by the vibrational modes that are most relevant to particle rearrangements, and the length scale is associated with the number of particles participating in a rearranging event.  Following Ref.~\cite{manning}, Fig.~3 suggests that the lowest frequency vibrational modes, particularly the lowest $\emph{N}_{m}=10$ of them, are most correlated with particle rearrangements; thus, the frequency of the 10th lowest frequency mode provides an estimate of the energy scale.  Similarly, the experiments show that the largest rearrangement observed contains $\emph{N}_{p}\approx 130$ particles; this determines the characteristic length scale. These choices for $\emph{N}_{m}$ and $\emph{N}_{p}$ are not arbitrary but have physical motivation; we also note that the results are quite insensitive to the particular values of $\emph{N}_{m}$ and $\emph{N}_{p}$, so that similar results are obtained for $\emph{N}_{m}=50$ and $\emph{N}_{p}=65$~\cite{supp}.  We limit our analysis to the high packing fraction regime in order to best match the conditions in simulations wherein only one rearrangement event occurs at a time.

The resulting soft spots are shown in comparison with the rearranging clusters in Fig.~4.  At high packing fractions, rearrangements reliably occur at soft spots, suggesting that soft spots are pre-existing structural defects wherein the material preferentially fails.  Note also that the average soft spot size is roughly 20 particles and the density of the soft spots is about 10\% of the total number of particles in the field of view at all  packing fractions studied.

Fig.~4 shows that while the soft spot analysis identifies a population of regions wherein the particles are likely to rearrange, it does not predict which of those regions will rearrange. In a thermal system such as ours, the precise location of the rearrangement is chosen 
stochastically, depending on the energy barriers encountered. This scenario differs fundamentally from the situation for systems sheared quasistatically at zero temperature, in which soft spots were previously identified \cite {manning}. In addition, we note that the simulation focused only on the first initial rearrangement event with no regard to possible avalanches or separate excitations at slightly later times. Experiments, on the other hand, are susceptible to avalanches and necessarily consider all rearrangement events, including secondary motions, that occur during the change of packing fraction. Nonetheless, Fig.~4 suggests that sufficiently far above the jamming transition, rearrangements tend to occur at soft spots even in our thermal system, consistent with the idea that soft spots are robust, intrinsic defects in disordered solids that control where local failure occurs under mechanical load. The experiments thus raise the possibility of using vibrational properties to identify regions susceptible to failure in packings of constituents ranging from nanoparticles to macroscopic grains, perhaps even including systems subject to natural events such as earthquakes and avalanches \cite{marone2002, douady1999}.

We thank N. Xu, C. Brito, S. Henkes for helpful discussions. This work was funded by DMR 0804881 (AGY), PENN-MRSEC DMR-0520020 (AGY, AJL, KC, WGE), NASA NNX08AO0G (AGY) and DOE DE-FG02-05ER46199 (AJL, WGE).


\end{document}